# An innovative *in silico* model of the oral mucosa reveals the impact of extracellular spaces on chemical permeation through epithelium

Sean M. Edwards[a,*], Amy L. Harding[b,*], Joseph A. Leedale[c], Steve D. Webb[c], Helen E. Colley[b], Craig Murdoch[b,1,§] and Rachel N. Bearon[a,d,1,§]

**Author affiliations:**

[a]Department of Mathematical Sciences, University of Liverpool, Liverpool L69 7ZL, UK

[b]School of Clinical Dentistry, University of Sheffield, Sheffield S10 2TA, UK

[c]Syngenta UK Limited, Jeallott's Hill International Research Centre, Bracknell, RG42 6EY, UK

[d]Department of Mathematics, King's College London, London, WC2R 2LS, UK

RNB's affiliation changes from Univeristy of Liverpool to Kings College London in December 2023.

[1]Correspondence; Email:  c.murdoch@sheffield.ac.uk or rachel.bearon@kcl.ac.uk

*Authors contributed equally to this work. §Joint senior and corresponding authors.

**Author Contributions:** SME, ALH, JAL, SDW, HEC, CM and RNB conceptualized research and designed methodology; SME and ALH performed investigative research, validation and formal data analysis; SME designed, implemented and tested computer code; CM, RNB and HEC secured funding and supervised the project; CM had overall responsibility for project administration; SME, ALH prepared the draft writing of the manuscript; all authors reviewed and edited the manuscript.

**Competing Interest Statement:** SME, ALH, HEC, CM and RNB declare no competing interests. JAL and SDW are employees of Syngenta UK Limited, although Syngenta did not fund any of this study.

**Classification:** PHYSICAL SCIENCES (Applied Mathematics) and BIOLOGICAL SCIENCES (Applied Biological Sciences)

**This PDF file includes:**
    Main Text
    Figures 1 to 4
**Supplementary file includes:**
    Supplementary Figures 1 to 5

**Abstract**

In pharmaceutical therapeutic design or toxicology, accurately predicting the permeation of chemicals through human epithelial tissues is crucial, where permeation is significantly influenced by the tissue's cellular architecture. Current mathematical models for multi-layered epithelium such as the oral mucosa only use simplistic 'bricks and mortar' geometries and therefore do not account for the complex cellular architecture of these tissues at the microscale level, such as the extensive plasma membrane convolutions that define the extracellular spaces between cells. Chemicals often permeate tissues via this paracellular route, meaning that permeation is underestimated.

To address this, measurements of human buccal mucosal tissue were conducted to ascertain the width and tortuosity of extracellular spaces across the epithelium. Using mechanistic mathematical modelling, we show that the convoluted geometry of extracellular spaces significantly impacts chemical permeation and that this can be approximated, provided that extracellular tortuosity is accounted for. We next developed an advanced physically-relevant *in silico* model of oral mucosal chemical permeation using partial differential equations, fitted to chemical permeation *in vitro* assays on tissue-engineered human oral mucosa. Tissue geometries were measured and captured *in silico,* and permeation examined and predicted for chemicals with different physicochemical properties. The effect of altering the extracellular space to mimic permeation enhancers was also assessed by perturbing the *in silico* model. This novel *in vitro-in silico* approach has the potential to expedite pharmaceutical innovation for testing oromucosal chemical permeation, providing a more accurate, physiologically-relevant model which can reduce animal testing with early screening based on chemical properties.

**Significance Statement**

There is increasing interest in how chemicals permeate through multi-layered epithelial tissue such as the oral mucosa for novel drug delivery purposes or to examine chemical-induced toxicity. *In silico* modelling can be used to aid research but current models do not account for the microscale structure of human tissue. Using a combination of *ex vivo-in vitro* measurements and experiments alongside mechanistic mathematical modelling, we show that inclusion of convoluted extracellular space and histologically-accurate cellular geometry within *in silico* models is imperative as these features impart significant impacts on chemical permeation. We propose an advanced *in vitro-in* *silico* approach to more accurately predict chemical permeation of oral mucosa, which could be adapted to other tissues.

**Introduction**

Buccal oral mucosa (the tissue lining the cheek) is now recognized as an attractive drug delivery site alternative to injection and oral administration, offering an array of benefits such as non-invasiveness, by-pass of first-pass metabolism by the liver, practicality of administration and increased patient compliance (1). The buccal mucosa is a particularly suitable site for drug delivery due to its anatomical and physiological characteristics. It is composed of a non-keratinized epithelium that is several-times more permeable than the skin and, along with its highly vascularized underlying connective tissue, promotes rapid chemical absorption that is integral for effective topical or systemic drug delivery (2). Currently, only a small number of chemicals are routinely delivered via the oral mucosa for systemic applications (e.g., glyceryl trinitrate for angina, fentanyl citrate for pain, prochlorperazine for nausea). However, the development of advanced polymer-based manufacturing technologies for mucoadhesive patches and other smart intraoral systems have marked a substantial step forward in this drug delivery domain (3, 4). It is projected that these emerging technologies will drive a significant expansion in the catalogue of pharmaceutical compounds available for administration through the oral mucosa (3, 5). In fact, successful clinical trials have already taken place for administration of patch-containing corticosteroids for the treatment of inflammatory oral lesions (6, 7). As the scope of compounds likely to be delivered via the mucosa expands, it is imperative for future drug screening and chemical toxicity testing that robust *in vitro-in silico* correlation models are established that accurately predict chemical performance *in vivo*. The integration of these models will expedite the translation of more realistic and relevant formulations from *in vitro* testing to full-scale commercial production, while simultaneously reducing the requirement for animal experiments during the pre-clinical phase.

*In silico* modelling of drug/chemical permeation through human tissues is highly advantageous for the development of novel pharmaceutics and toxicological screening. The current *in silico* predictive approaches for epithelial tissues are empirical models that directly fit permeation rates to the biophysical properties of a chemical (8, 9), compartmental analysis with well-mixed homogeneous tissue layers (10-12) or approximate diffusive models that are based on 'brick and mortar' cellular geometries (13, 14). These models do not take into account the real tissue structure of the epithelium and ignore the spaces that exist between cells (extracellular spaces), which then underrepresents chemical permeation around cells via the paracellular route. Alternatively, the microscale geometry of a tissue can be included by using partial differential equations (PDEs). Such models can elucidate the impact of microscale factors such as cellular geometry, permeation of cell membranes and cellular metabolism, which contribute to permeation differently for different chemicals. A cell-based approach of this type was recently established in liver spheroids (15), which allowed for extracellular and intracellular predictions and consideration of paracellular permeation.

In this study, the microscale structure of the oral epithelium was investigated to understand the mechanisms behind drug/chemical permeation through tissue. For the first time, we quantify the extracellular space on a subcellular scale and measure its tortuosity in both human normal oral mucosa (NOM) and tissue-engineered normal oral mucosa (TENOM). We use these data to inform an innovative *in silico* model of chemical permeation through the oral mucosa, which incorporates the histological structure of the epithelium, containing discrete cells separated by extracellular space. We further include cellular tortuosity and demonstrate its importance within the mucosa upon validating the model with an exemplar chemical, oxymetazoline HCL. The data generated from *in vitro* investigations of chemical permeation, parameterized and extrapolated to an *in silico* model, provides a more comprehensive understanding of the subcellular structural impact for oral mucosal drug delivery and chemical permeation, the principles of which could be extrapolated to assess chemical permeation *in vivo* or in other epithelial tissues.

**Results**

***Oral epithelium displays a convoluted extracellular space topology***
The oral mucosal epithelium is divided into three distinct layers based on cell differentiation. Basal cells attach to the basement membrane of the connective tissue (lamina propria) and display stem cell-like qualities. Moving apically, the basal cells differentiate into the spinosum layer, where cells are larger in size, and finally into the superficial layer, where the terminally differentiated cells are thin and flat (Fig. 1A). Transmission electron micrographs show that the extracellular spaces between epithelial cells are convoluted and inhomogeneous throughout the epithelial strata (Fig. 1B-D); whilst this has been observed in other tissues (e.g., skin; (16)) this is the first time it has been measured and characterized in this manner for oral epithelium. In NOM, the superficial and spinous strata display a significantly narrower extracellular space with convolutions of less amplitude and wavelength than the basal cells (Fig. 1E, $p < 0.01$), that may restrict paracellular chemical permeation. Corresponding results of extracellular space convolutions for TENOM are provided in Fig. S1.

The extracellular space is convoluted in every stratum of oral epithelium, increasing cellular surface area and lengthening the extracellular path around cells. From the basal layer to the superficial layer, the amplitude and wavelength of the convoluted spaces decrease, which individually affect tortuosity in opposing directions. The net effect of this is that the subcellular tortuosity $\tau_S$, (i.e., the extracellular pathlength divided by the straight pathlength (Fig. S2C)), is similar throughout the epithelium with $\tau_S > 1$ (Fig. 1F).

The height and breadth of cells in each stratum of the oral epithelium were also measured (Fig. 1G) along with the thickness of each stratum (Fig. 1H). Corresponding measurements for TENOM are provided in Fig. S1. As cells differentiate from the basal layer to the superficial layer their breadth increases resulting in cellular elongation and characteristic flattening. This corresponds to increased tortuosity on a cellular scale as cells differentiate apically ('cellular tortuosity'), and the overall oral mucosal tortuosity $\tau$ (the product of subcellular and cellular tortuosities) also follows this trend. A comparison between NOM and TENOM show that the basal and superficial strata display similar characteristics while the spinous strata thicknesses is significantly increased in NOM compared to TENOM ($p<0.0001$) (Figs. 1H and S1H).

### *Convoluted extracellular spaces affect chemical permeation dynamics*

A physically-relevant *in silico* model of a basal cell monolayer, with discrete cells and extracellular spaces, was used to investigate the impact of convoluted extracellular space on chemical permeation into and around cells. A schematic of this model is shown in Fig. S2A, with convoluted geometry based on experimental measurements from NOM (Fig. 1E-G). A constant supply of chemical is applied at the apical cell surface with concentration $C_0$ and below the cells a perfect sink ($C = 0$) mimics chemical delivery to the lamina propria or vascular uptake. This approximation is good for basal cells which lie near to the lamina propria, but the same system can also be used to study the effect of convoluted gaps in the superficial and spinous strata. Here, the breadth and height of cells along with convoluted space measurements are adjusted appropriately. A test chemical is supplied apically at time $t = 0$s with extracellular and intracellular diffusion coefficients $D_E = D_I = 7.5 \times 10^{-10}$m$^2$s$^{-1}$. In practice the diffusion coefficient depends on the physicochemical properties of a chemical but previous analysis of the diffusion coefficients of over three hundred small molecule chemicals has suggested that this is reasonable as a representative value (15).

Including the convoluted extracellular space makes substantial and predictable changes to chemical permeation dynamics, and the *in silico* model can also predict variability across different epithelial strata. This is illustrated in Fig. 2 for three example chemicals with differing ability to permeate the cell membrane (coefficient $Q$, dependent on the physicochemical properties of a chemical), mimicking their dependence on lipophilicity with: lipophilic (highly membrane permeating chemicals; Fig. 2A); moderately lipophobic (moderately membrane permeating chemicals; Fig. 2B) and perfectly lipophobic (membrane impermeable chemicals, Fig. 2C). As a constant supply of chemical is applied the system reaches a steady-state (Fig. 2Ai–Ci). Average concentrations over time are deduced in the extracellular space between cells (Fig. 2Aii–Cii) and intracellularly, within the cell (Fig. 2Aiii–Biii).

The lipophilicity of a chemical alters its dynamics inside and outside cells. For highly membrane-permeating (lipophilic) chemicals (Fig. 2Ai, $Q = 10^{-3}$ ms$^{-1}$), the average chemical concentrations in the extracellular space (Fig. 2Aii) and intracellularly (Fig. 2Aiii) are identical, indicating that the geometry of the extracellular space has little impact on permeation. In this lipophilic case, permeation through the monolayer is governed by diffusion, with diffusion time increasing monotonically with cell height $h$ (Fig. 1G). However, moderately membrane-permeating (moderately lipophobic) chemicals have a different steady-state profile intra and extracellularly (Fig. 2Bi, $Q = 10^{-5}$ ms$^{-1}$). The time to reach steady-state is significantly slowed in both regions (Figs. 2Bii, 2Biii), with extracellular space and intracellular concentrations remaining coupled. In contrast to lipophilic chemicals, the time to reach steady-state is no longer in order of cell height, indicating the influence of other geometric factors including subcellular tortuosity or cell area. Impermeable chemicals ($Q = 0$), are restricted to the extracellular space and chemical permeation dynamics are dominated by diffusion along the convoluted paracellular route (Fig. 2Ci). The subcellular tortuosity of the gap increases the length-scale for diffusion relative to lipophilic chemicals, thus increasing the time to reach steady-state in the extracellular space (Fig. 2Cii). However, this process is non-linear as the timescale to reach steady-state is faster than moderately lipophobic chemicals.

Further analysis of highly membrane permeating lipophobic chemicals ($Q < 10^{-6}$ ms$^{-1}$) suggests that they first permeate the extracellular space before then permeating into cells (Fig. S3). The steady-state concentration within cells becomes more uniform as the chemicals become more lipophobic.

***Reduced extracellular diffusion approximates effect of convoluted topology in silico***
When compared against a straight extracellular space model in which the convoluted subcellular geometry has been removed, permeation through the monolayer is slower for the convoluted gaps than for the straight gaps (Fig. S4). However, an equivalence can be found if we introduce a slower effective diffusion coefficient for the straight gaps,

$$D_E^{straight} = \frac{D_E}{\tau_S^2}. \tag{1}$$

By appropriately slowing extracellular diffusion in the 'straight' extracellular space model (Fig. S2B) to incorporate the subcellular tortuosity of spaces (Eqn. 1), the permeation dynamics of the physically relevant 'convoluted' cell monolayer are captured, particularly for spinous and superficial cells. This approximation can be applied in more sophisticated *in silico* models involving multiple cells for which the computational cost of incorporating the subcellular geometry would be prohibitive. If subcellular tortuosity is ignored there will be a consistent underestimation of the permeation timescales (Fig. S4).

***An in vitro - in silico approach to study tortuous impact on oral mucosal permeation***

We next developed an *in silico* model to capture the impact of extracellular convolutions on chemical permeation through TENOM. Data from *in vitro* epithelial models rather than NOM are used here because NOM is not available for *in vitro* permeation experiments due to unavailability of human tissue and that animal oral mucosa is significantly different to human. TENOM is the best alternative in terms of morphology to NOM. By incorporating measurements of extracellular space widths (Fig. S1E), cell geometry (Fig. S1G), epithelial layer thickness (Fig. S1H), and nuclei coordinates (Fig. 3B), the *in silico* model is geometrically comparable to the cellular histological structure of TENOM (Fig 3A-D). Cell boundaries are deduced with a modified Voronoi tessellation in which the elongation of cells from the basal to the superficial layer is accounted for (Fig. 3C). Extracellular spaces have width $w$ (Fig. 3D) equivalent to the measurements obtained in each epithelium strata.

A schematic of the mathematical model is provided in Fig. 3E. Chemical permeation through the spinosum and basal layers is modelled with diffusion in the extracellular spaces and intracellular regions (coefficients $D_E$ and $D_I$) and the ability of chemicals to permeate cell membranes (coefficient $Q$). The extracellular spaces in this model are straight, but incorporate the established convoluted geometry, where extracellular diffusion is reduced by subcellular tortuosity $\tau_S$ (Eqn. 1). A chemical is applied in an application layer and included are superficial and lamina propria layers. At the base of the lamina propria, perfect sink conditions mimic a relatively large volume of receptive medium used in the tissue culture conditions to produce TENOM (Fig. 4A).

Oxymetazoline hydrochloride (HCl) (logP = 3.03), a small molecule vasoconstrictor that acts on blood vessels in the lamina propria, is topically applied in solution [3.4 mM] to the surface of TENOM (Fig. 4A) and the percentage permeation determined by measuring the concentration of oxymetazoline HCl in the receptive medium (phosphate-buffered saline (PBS), pH 7.4) over time (Fig. 4B) by high-liquid performance chromatography. After four hours 53.6 ± 15.7% of oxymetazoline HCL had permeated the TENOM, rising to 77.8 ± 20.0% by six hours. The *in silico* model was then fitted to this data. With diffusion coefficients in each layer already established, fitting was undertaken using Newton iteration in the chemical lipophilicity parameter $Q$ and for oxymetazoline HCl we deduce $Q = 4.7 \times 10^{-7}$ ms$^{-1}$. Predicted *in silico* oxymetazoline HCl concentrations throughout the entire TENOM epithelium are given in Fig. 4C over 360 minutes, alongside magnifications of the epithelium (Fig. 4D) and the extracellular space (Fig. 4E). As oxymetazoline HCl is relatively lipophobic, the concentration profile within individual cells is nearly uniform, leading to a step-like concentration profile from the basal layer to the superficial layer.

In Fig. 4F the extracellular space is adjusted to mimic permeation enhancing chemicals that can alter the geometry of the convolutions between cells. Widening the extracellular space alone has minimal effect on overall permeation (50% permeated at 3.65 hours); however, straightened gaps display enhanced permeation (50% at 2.53 hours), which is substantially increased with extracellular space widening (50% at 1.60 hours). These effects of permeation enhancers on the convoluted extracellular space are important for lipophobic drugs like oxymetazoline HCl. Indeed, the rate of tissue permeation of highly lipophobic chemicals that are membrane impermeable will be highly dependent on extracellular space. In contrast, lipophilic chemicals are predicted to permeate much more quickly, and permeation is unaffected by tissue geometry in this case (Fig. 4G; 50% at 0.31 hours for hypothetical chemical X with the same properties as oxymetazoline HCl except lipophilic).

**Discussion**

Administration of drugs/chemicals via the oral buccal mucosa provides several clinical and pharmacokinetic advantages that can positively impact patient pharmacotherapy, including avoiding stomach acid, circumventing hepatic first-pass clearance, and its suitability for chemically unstable and sensitive substances (17, 18). Rodents are the animal of choice for many *in vivo* drug delivery and chemical toxicity studies but their usefulness here is highly questionable as the structure of their oral mucosa is vastly different to humans. Rodent oral mucosa is highly keratinized and impermeable at the surface, resembling skin rather than the non-keratinized, permeable epithelium that is present in the human oral cavity (19, 20). Animal experimentation has some other obvious disadvantages such as costly experimentation and ethical considerations. However, post-surgical *ex vivo* human oral mucosa is minimally available for laboratory experimentation (20). Alternatively, permeation assays using TENOM have the potential to reduce animal need and expedite the screening process for novel chemical compounds. The TENOM used in this study display the fundamental tissue architecture and cellular characteristics of NOM, and have been validated in terms of cell differentiation status, cell marker expression and chemical permeability (21, 22). However, a difference does occur with the relative size of the spinous layer. With the future aim to extrapolate *in vitro* drug permeation results to *in vivo*, this difference can be factored in to the *in silico* model that incorporates all the other measured tissue geometries.

*In silico* models are now firmly recognised as an effective approach for experimental extrapolation and outcome prediction of tissue permeation (23) but their transferability to clinical or regulatory use is highly dependent on the data used and the approximations made. To date, the information available on the permeability of chemicals across epithelia largely come from skin, whereby several mathematical models have been developed using skin-specific data to predict transdermal chemical permeability (24). Indeed, mathematical models of the oral mucosa are very sparse (8, 9, 12) . The buccal mucosa displays several distinct features (non-keratinized, thicker epithelium with increased permeability), making it dissimilar to skin. Since these features predominantly determine the rate-limiting step for transepithelial drug delivery, mathematical models developed to predict drug delivery across the skin are not entirely applicable to that of the oral mucosa, warranting separate investigations of oral mucosal-specific parameters that has driven the necessity for this research.

Like many other tissues (e.g., skin (16), liver (25), brain (26), vaginal mucosa (27)), buccal mucosa exhibits convoluted extracellular spaces, and since many chemicals permeate tissue via this paracellular route, these spaces may significantly impact permeation. Whilst the impact of cell-scale tortuosity due to differing cell sizes and shapes has been considered before (15), subcellular convoluted geometries have been largely ignored. For the first time, the influence of convoluted extracellular space on drug permeation has been elucidated. Highly lipophilic, membrane impermeable chemicals that are restricted to the paracellular route are significantly slowed as they pass through the convoluted extracellular space. However, increased cellular surface areas improve transcellular permeation for moderately lipophobic membrane-permeable chemicals. These contrasting effects depend on the physicochemical properties of the administered chemical and accounting for these is particularly important for lipophobic chemicals. We have shown that our *in silico* model can account for the convoluted geometries by appropriately reducing extracellular diffusion, making a significant advancement in this area of research.

This study introduces an innovative whole-epithelium *in silico* model describing the marked impact that extracellular spaces of the oral mucosa have on drug permeation. Employing buccal tissue and cell geometry measurements, our model surpasses existing alternatives in its sophistication and accuracy. Crucially, by using partial differential equations, physical chemical permeation dynamics are included, such as extracellular and intracellular diffusion and permeation of cell membranes, along with real cell-shapes, cell-sizes and convoluted extracellular spaces; a significant improvement on previous models. These parameters can be directly perturbed in the model. If this tissue geometry is ignored, the time of oxymetazoline HCl to reach 50% permeation is underpredicted by 30%. Prediction of permeation for chemicals with different physiochemical properties, with lipophilicity being a key determining factor, may be even more inaccurate. This discrepancy in predictive power may have significant implications for investigators aiming to understand the kinetics of drug delivery to the oral mucosa, drug availability within epithelial tissue or maintaining clinically-relevant levels of drugs over specific periods of time.

Permeation can also be improved if the extracellular space is enlarged, for example with tissue permeation enhancers (28, 29), the modelling of which is crucial if chemical permeation through tissues is to be improved to expedite systemic delivery. Our model correctly predicts that lipophilic (diffusion-limited) chemicals are minimally affected by cellular geometry, while the opposite is true for lipophobic chemicals. This feature will allow investigators to determine which permeation enhancers to select for a given chemical to fine tune its systemic delivery.

Our combined *in vitro* – *in silico* approach has the potential to significantly improve oral pharmacokinetics and advance buccal-delivered therapies. The next step is to develop a comparative *in silico* model of *in vivo* tissue, in which chemicals pass to the bloodstream, parameterized by further *in vitro* permeation studies. Our *in silico* model could then be linked to whole body physiologically-based pharmacokinetic (PBPK) models, allowing prediction of the distribution of a chemical throughout the whole body. The methodology developed here also has the potential to impact wider drug distribution modelling, if adopted for other tissues and modalities, improving predictability and reducing the need for animal testing.

**Materials and Methods**

All reagents were purchased from Merck (Gillingham, United Kingdom) and used as per the manufacturers' instructions unless otherwise stated.

*Ex vivo oral tissue and tissue-engineered normal oral mucosa (TENOM) models*

TENOM were constructed using immortalized oral keratinocytes (FNB6) and NOF as previously described (21). *Ex vivo* buccal tissue was sourced from a combination of archived formalin-fixed, paraffin-embedded oral mucosal tissue held by the Unit of Oral Maxillofacial Pathology, Sheffield Teaching Hospitals NHS Foundation Trust (Ethical approval number 07/H1309/105) and freshly excised buccal tissue from healthy patient volunteers undergoing routine maxillofacial surgery with written, informed consent (Ethical approval number 09/H1308/66).

*Histological image analysis*

NOM biopsies and TENOM were fixed with 10% (v/v) neutral-buffered formalin; alcohol-processed, paraffin wax embedded, 5 μm sections cut using a microtome and sections stained with Haematoxylin & Eosin (Epredia, UK). Slides were mounted with distyrene-polystrene xylene and imaged by light microscopy. Images were digitised using an Aperio Slide Scanner (Leica) and extracellular space widths measured using the associated Aperio ImageScope software (Leica Biosystems, UK) or ImageJ software (NIH).

*Extracellular space analysis using transmission electron microscopy (TEM)*

NOM biopsies and TENOM were fixed with 3% glutaraldehyde diluted in 0.1 M cacodylate buffer (pH 7.4), for 2 hours at 4°C before rinsing twice in 0.1 M phosphate buffer (pH 7.4) at 4°C. Samples were subjected to post-fixation, dehydration, and resin embedding as described in (22). Sections were examined for extracellular spaces using an FEI Tecnai TEM at an accelerating voltage of 80 kV and images taken using a Gatan digital camera. Post-image measurements were conducted on the convoluted extracellular spaces by assuming that they are locally sinusoidal with width $w$, amplitude $A$ and wavelength $\lambda$. Measurements were collected for basal, spinosum and superficial epithelial strata. Extracellular space widths were measured using ImageJ software (NIH).

*Permeation assay*

TENOM were maintained at 37°C, 5% $CO_2$ in a humidified incubator and 50 μL oxymetazoline HCL [3.4 mM] applied topically to the apical surface of the epithelium for up to 24 hours. After chemical exposure, the receptor PBS (pH 7.4) solution in the basolateral chamber was sampled (200 μL) in a time-dependent manner (for up to 360 minutes), and the volume replaced with fresh pre-warmed

PBS (pH 7.4) to maintain experimental conditions. The concentration of oxymetazoline HCl permeated through the model was determined using high performance liquid chromatography (HPLC). A schematic of this assay design can be seen in Fig. 4A.

*High-performance liquid chromatography*

Sample analysis was performed with a Shimadzu Prominence HPLC instrument, using a 250 mm × 4.6 mm column and a mobile phase composed of acetonitrile (MeCN (37.5%)/(MeOH (37.5%)/ 10 mM $NH_4(CO_3)_2$ (25%) for 10 minutes with a flow rate of 1 mL min$^{-1}$. Emission was detected at 308 nm and the concentration of sample calculated from standard calibration curves [1-50 μg/mL; 3.37-168 $\mu$M] (n=3 independent experiments).

*Data analysis*

All data are presented as mean ± SD, with all experimental repeats clearly stated. Parametric data was analysed by ordinary one-way ANOVA with Tukey's post-hoc test for multiple group comparisons. Statistical analysis was performed using GraphPad prism, version 9.0 (GraphPad Software, San Diego, CA) and significance assumed if p<0.05.

*Tortuosity calculations*

Subcellular tortuosity $\tau_S$ of convoluted gaps is the sinusoidal pathlength compared to the wavelength (Fig. 1F),

$$\tau_S = \frac{1}{\lambda}\int_0^\lambda \left[1 + \left(\frac{2\pi A}{\lambda}\cos\left(\frac{2\pi \tilde{x}}{\lambda}\right)\right)^2\right]^{1/2} d\tilde{x}. \tag{2}$$

Cellular tortuosity $\tau_C = (h+b)/h$ is approximated from the height $h$ and breadth $b$ of cells (Fig. 1G). The overall tortuosity of the tissue $\tau = \tau_C \tau_S$ is a combination of these contributions.

*In silico model of chemical permeation through an oral epithelial monolayer*

The *in silico* monolayer model (Fig. S2A) has rectangular cells mimicking a buccal basal layer; cell breadth $b$ (aligned with $x$) and height $h$ (aligned with $y$) match NOM experimental measurements. The domain has a vertical extracellular gap (width $w$) and horizontal gaps above and below the cell-layer (widths $w/2$). Extracellular spaces between cells are sinusoidal, approximating experimental measurements (Fig. 1E). Chemicals diffuse through the extracellular and intracellular regions (coefficients $D_E, D_I$) with concentrations ($C_E, C_I$),

$$\frac{\partial C_E}{\partial t} = D_E \nabla^2 C_E, \qquad \frac{\partial C_I}{\partial t} = D_I \nabla^2 C_I. \tag{3}$$

Passive transport of chemical through cell boundaries is modelled with a permeability coefficient $Q$,

$$D_E \nabla C_E \cdot n = D_I \nabla C_I \cdot n = Q(C_E - C_I), \tag{4}$$

with flux conserved; a full discussion is given in (15). A constant supply of chemical with concentration $C_0$ is applied at the top and a perfect sink mimics vascular uptake at the bottom,

$$C_E(y=0) = C_E(y=H+w) - C_0 = 0. \tag{5}$$

Each cell has the same concentration profile with symmetry on the side boundaries,

$$\frac{\partial C_E}{\partial x}(x=0) = \frac{\partial C_I}{\partial x}(x=0) = \frac{\partial C_E}{\partial x}(x=W+w) = \frac{\partial C_I}{\partial x}(x=W+w) = 0. \tag{6}$$

### *In silico model of chemical permeation through tissue-engineered normal oral mucosa*

Cell nuclei coordinates inferred from stained tissue sections (Fig. 3B) were used to generate cells via Voronoi tessellation (Fig. 3C, see (30)). Cell elongation from the basal to the superficial layer (Fig. 1) is captured by adjusting the metric such that the distance $d$ between two points $p_0(x_0, y_0)$ and $p_1(x_1, y_1)$ is,

$$d(p_0, p_1) = [(x_1 - x_0)^2 + \beta^2(y_1 - y_0)^2]^{1/2}, \tag{7}$$

with elongation parameter $\beta = h/b$ ($\beta = 1$ is Euclidean and $\beta > 1$ cells are elongated in the $x$ coordinate). Elongation $\beta$ is assumed to increase linearly throughout the epithelium. To the authors knowledge this is the first time elongated Voronoi tessellation has been applied *in silico* (cell elongation is common in other tissues (31)); though this approach has been discussed theoretically by (32). Extracellular spaces are imposed with the polygon offset algorithm with gap widths $w$. There is no loss by evaporation or to the mouth and diffusion in each layer is modelled similarly to (Eqn. 3). Along the boundaries between layers flux is conserved; cell membrane permeation is modelled by (Eqn. 4).

### *Solving the in silico models numerically*

Diffusion equations (Eqn. 3) are solved in each region subject to (Eqns. 4-6) with a finite-element scheme and time-stepping with the Crank-Nicolson method. Three initial implicit-Euler steps are taken to accommodate discontinuous initial conditions. The linear solve phase is undertaken in C++ with the parallel sparse solver MUMPS (33) under the framework of PETSc (34). Solutions are ensured to be independent of discretization. Average concentrations are computed by integrating over a relevant region and dividing by its area.

*Parameterizing the in silico model to capture permeation of oxymetazoline HCl*

Diffusion in the application $D_A$ and lamina propria $D_{LP}$ layers was approximated to water $D_W$, determined with the Stokes-Einstein equation (35) for a given chemical,

$$D_A = D_{LP} = D_E \tau_S^2 = D_W = \frac{k_B T}{6\pi \eta_W R}; \tag{8}$$

with Boltzmann constant $k_B = 1.38065 \times 10^{-23}$ m² kg s⁻² K⁻¹, physiological temperature $T = 310.15 K$, dynamic viscosity of water $\eta_W = 6.913 \times 10^{-4}$ kg m⁻¹ s⁻¹, and spherical chemical radius $R$. Extracellular diffusion $D_E$ is reduced to account for subcellular convoluted tortuosity ($\tau_S$=3.576 for TENOM spinous layer). The chemical radius is inferred with,

$$R = \left(\frac{3M}{4\pi N_A \rho}\right)^{1/3}, \tag{9}$$

Avagadro's constant $N_A = 6.0221 \times 10^{23}$, which for oxymetazoline HCl (molecular mass $M = 296.84$, density $\rho = 1.1$ g cm⁻³) implies $R = 4.75 \times 10^{-10}$ m and $D_W = 6.92 \times 10^{-10}$ m²s⁻¹. Diffusion in the intracellular $D_I$ and superficial $D_S$ regions is reduced through the presence of organelles, which depends on the size of the chemical. We adopt the model suggested by Kwapiszewska *et al.*, (36) for human epithelial cell lines,

$$D_I = D_S = \frac{D_W}{1.3} \exp\left[\left(\frac{(4.6 \times 10^{-9})^2}{20^2} + \frac{(4.6 \times 10^{-9})^2}{R^2}\right)^{-0.285}\right]^{-1}; \tag{10}$$

for oxymetazoline HCl this predicts $D_I = 4.05 \times 10^{-10}$ m²s⁻¹. To fit the predicted percentage permeated $P(Q,t)$ to *in vitro* permeation data, the cell membrane permeability $Q$ is Newton iterated until the difference between the two curves is minimised in the L2 norm.

**Acknowledgments**

The authors would like to thank Dr Rob Bolt (Sheffield Teaching Hospitals NHS Foundation Trust) for collection of human tissue, Prof Ali Khurram (University of Sheffield) for sourcing archival paraffin wax embedded human oral mucosal tissue. We also appreciate the support of Mr Chris Hill of the Cryo-Electron Microscopy Facility, University of Sheffield for help with transmission electron microscopy. This study was funded by the National Centre for the Replacement, Reduction and Refinement of Animals in Research (NC3Rs) UK, grant number NC/W001160-1.

**Figures**

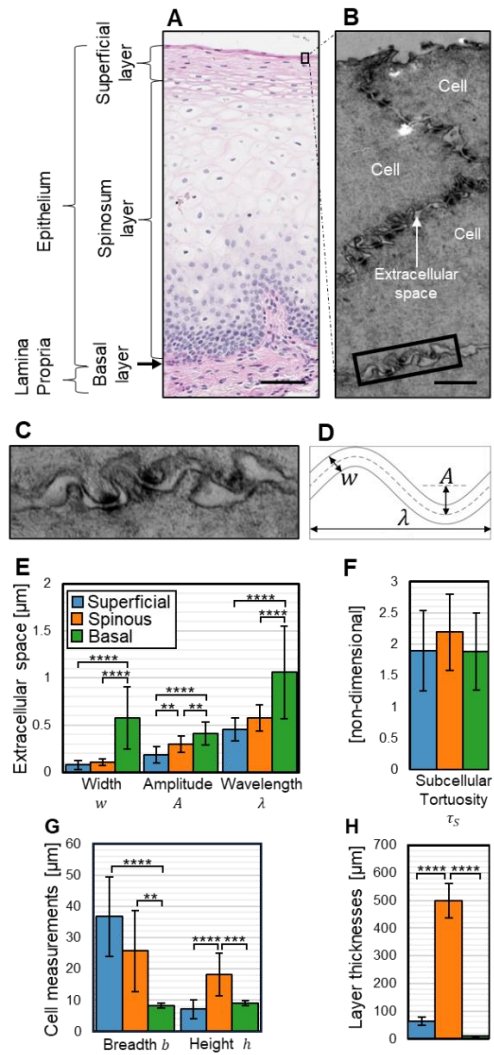

**Figure 1. Quantification of extracellular space convolutions and tissue geometry on the subcellular scale in the oral mucosa.** (A) Histological section of a normal oral mucosal biopsy displaying distinct cell layers, scale bar = 100 µm. (B) Transmission electron micrograph images of cells in the superficial layer displaying convoluted extracellular spaces between cells on a subcellular scale, scale bar = 1 µm. (C) Magnified image of convoluted extracellular spaces in B reveals a quasi-periodic profile of extracellular spaces. (D) Schematic of the extracellular space convolutions, assuming a sinusoidal profile with space width $w$, amplitude $A$ and wavelength $\lambda$. (E) Measurements of extracellular space width, amplitudes and wavelengths and (F) subcellular tortuosity in each oral epithelial strata of normal oral mucosa (NOM), legend in (E) applies to (F,G,H). (G) Cell breadth and height, and (H) cell layer thicknesses measured in each strata. Data is expressed as mean ± S.D., **p <0.01, ***p <0.001 and ****p <0.0001 as analyzed by ordinary One-way ANOVA with Tukey's correction (n = 3 independent biopsy with at least 10 measurements taken from each image).

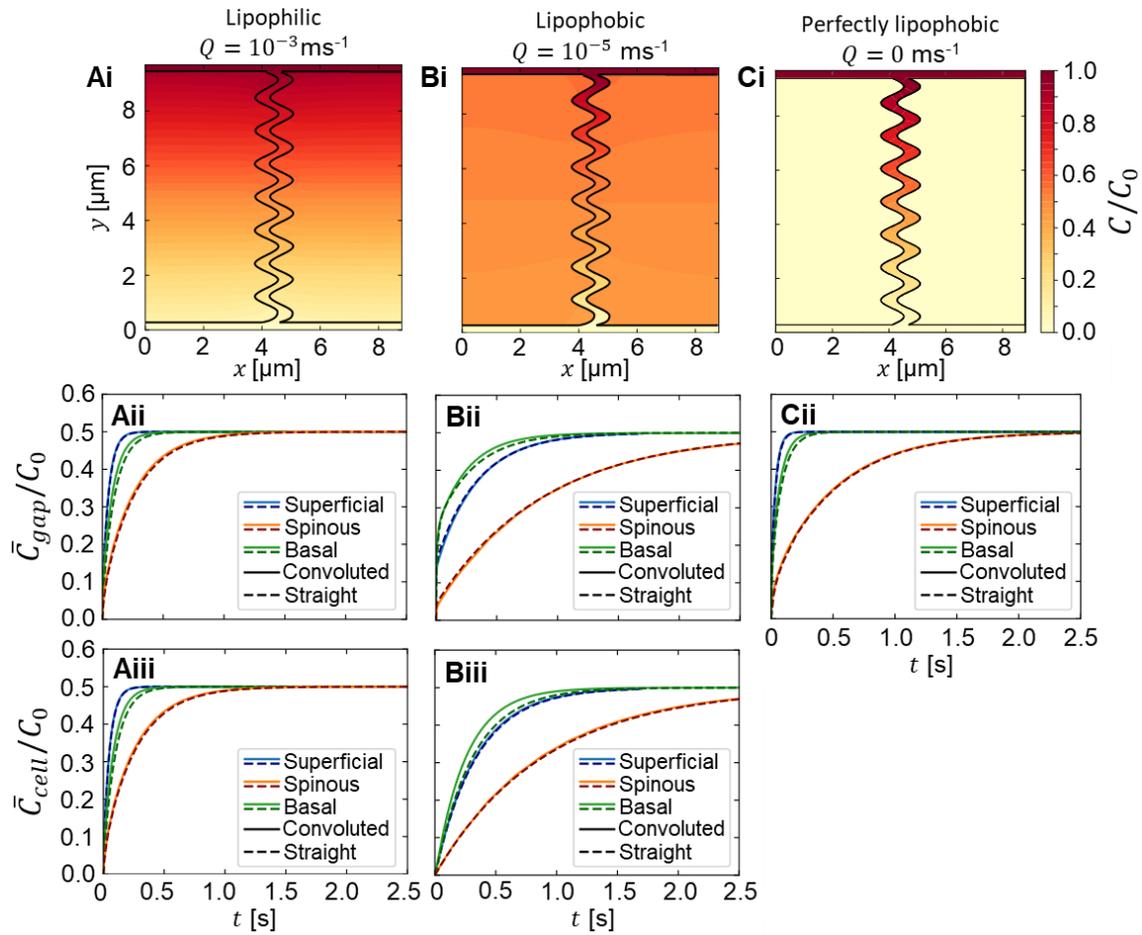

**Figure 2. Chemical permeation through a monolayer of oral epithelial cells modelled *in silico*.** A constant supply of chemical (concentration $C_0$) is passively transported through permeable membranes. Vascular uptake at the bottom is modelled by a perfect sink. Chemicals diffuse in the extra- and intracellular regions (coefficients $D_E = D_I = 7.5 \times 10^{-10}$ m²s⁻¹). The physiochemical property of each model chemical $Q$ is varied: (A) lipophilic (highly membrane permeable) $Q = 10^{-3}$ ms⁻¹; (B) moderately lipophobic (moderately membrane permeable) $Q = 10^{-5}$ ms⁻¹; and (C) perfectly lipophobic (impermeable) $Q = 0$ ms⁻¹. The concentration profile at steady state is presented (Ai, Bi, Ci) alongside average concentrations over time in the central extracellular space (Aii, Bii, Cii) and intracellularly (Aiii, Biii). Each type of epithelial cell strata is modelled (superficial, spinous, basal) in the full convoluted model (solid line) and results are compared with a straight approximation (dashed line) that has reduced extracellular diffusion (equation 1).

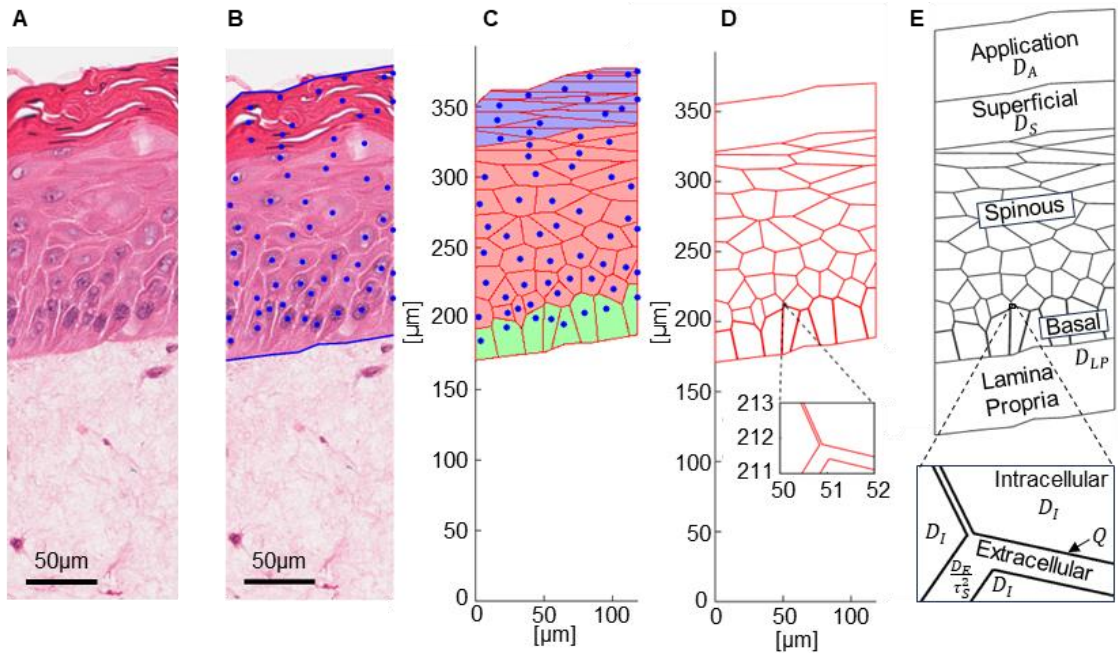

**Figure 3. *In silico* model development of chemical permeation through tissue-engineered normal oral mucosa.** (A) Haematoxylum and eosin (H&E) stained tissue section of TENOM showing basal, spinous and superficial epithelial strata on top of the lamina propria. (B) Nuclei coordinates (blue dots), inferred from histological H&E staining of TENOM, are used to construct cell boundaries using (C) Voronoi tessellation where epithelial strata are identified as basal (green), spinous (red) and superficial (blue). (D) Extracellular spaces are imposed on the basal and spinous strata and the superficial layer is isolated. (E) Chemical permeation is modelled with diffusion (coefficient $D$) and chemical lipophilicity (coefficient $Q$). Diffusion in each region can differ: application $D_A$, superficial $D_S$, extracellular $D_E$, intracellular $D_I$, and lamina propria $D_{LP}$ (TENOM lamina propria with thickness 942 ± 132µm). Extracellular diffusion is slowed to account for subcellular tortuosity $\tau_S$.

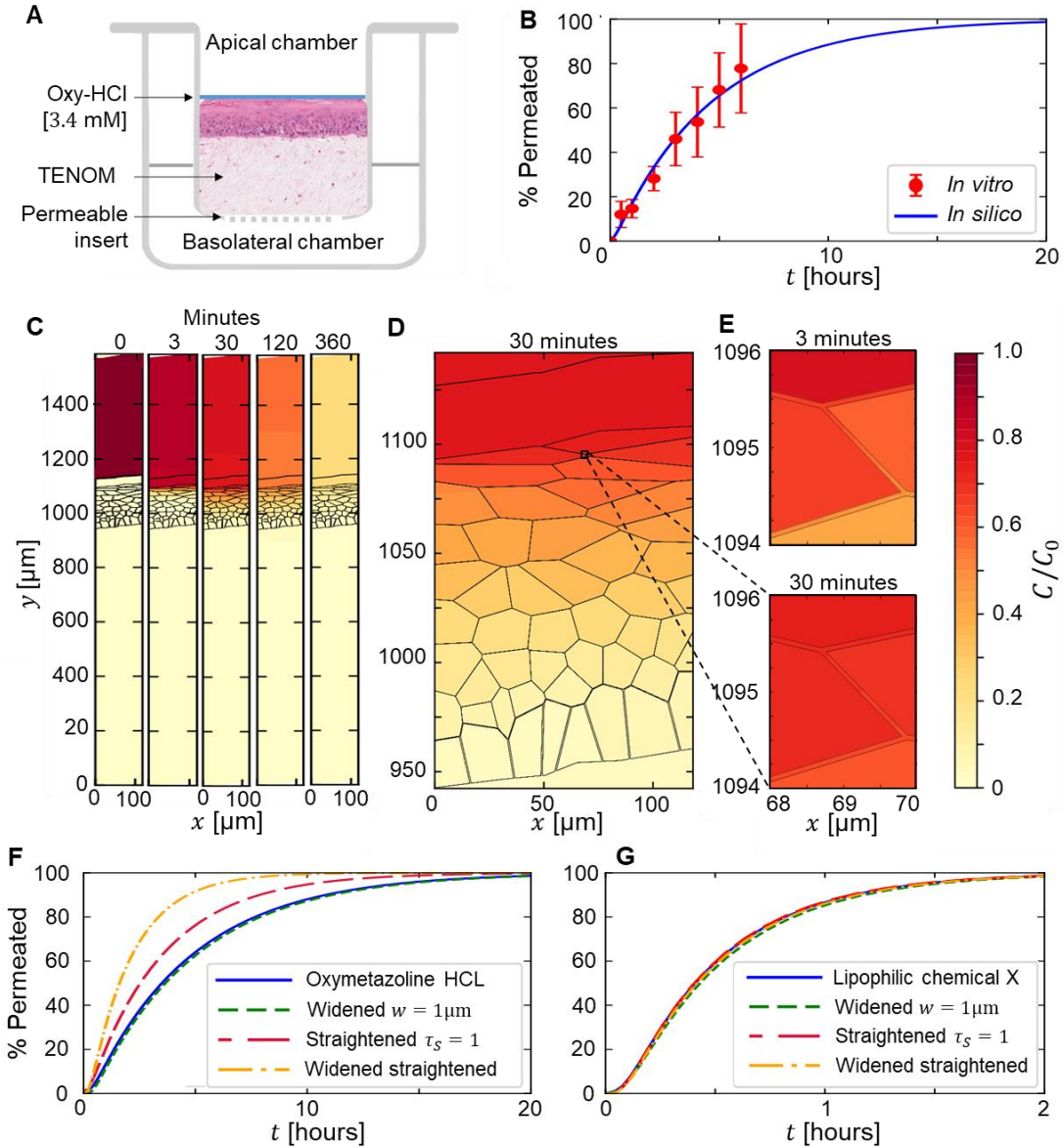

**Figure 4. Permeation of oxymetazoline HCl through tissue-engineered normal oral mucosa modelled *in vitro* and *in silico*.** (A) *In vitro* permeation assay with 50 μl of oxymetazoline HCl (concentration $C_0 = 3.4$ mM) applied to a 113.1 mm² transwell surface (2 mL PBS in basolateral chamber as sink). (B) *in vitro* permeation data (mean ± SD) is captured *in silico* by fitting chemical lipophilicity ($Q = 4.7 \times 10^{-7}$ ms⁻¹). (C) Predicted oxymetazoline HCl concentrations $C$ over 360 minutes by the *in silico* model in the whole tissue. (D) Magnification of the entire epithelium strata and (E) magnification of the extracellular spaces in the upper spinous strata. (F) Perturbation of microscale parameters affects whole-epithelium permeation timescales. (G) Chemical X, with increased lipophilic physiochemical properties ($Q = 10^{-3}$ ms⁻¹) compared to oxymetazoline HCl, permeates the mucosa much more quickly irrespective of tissue geometry.

**Supporting Information**

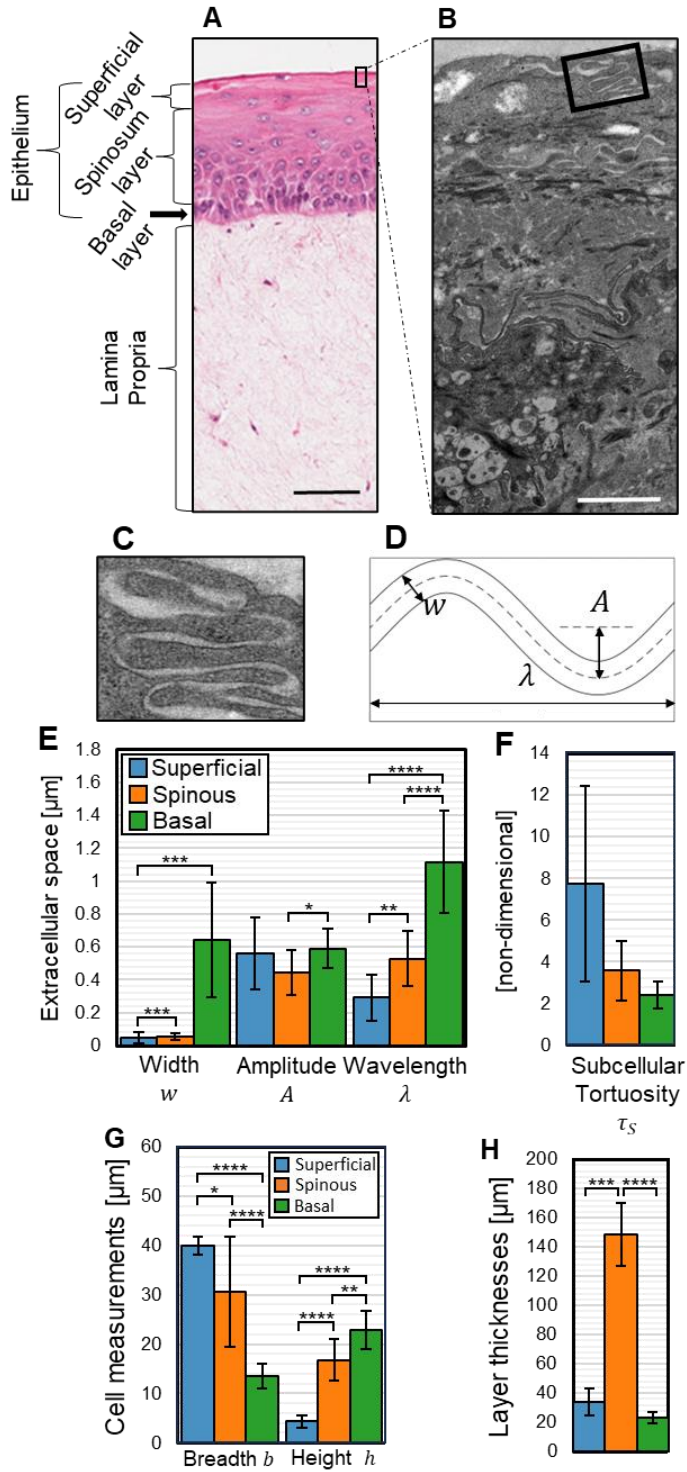

**Supplemental figure 1. Quantification of extracellular space convolutions and tissue geometry on the subcellular scale in tissue-engineered normal oral mucosa (TENOM).** (A) Histological section of TENOM displaying distinct cell layers, scale bar = 50 µm. (B) Transmission electron micrograph images of the superficial layer displaying convoluted extracellular spaces between cells on a subcellular scale, scale bar = 1 µm. (C) Magnified image of convoluted extracellular spaces in B reveals a quasi-periodic profile of extracellular spaces. (D) Schematic of the extracellular space convolutions, assuming a sinusoidal profile with space width $w$, amplitude $A$ and wavelength $\lambda$. (E) Measurements of extracellular space width, amplitudes and wavelengths and (F) subcellular tortuosity in each oral epithelial strata, legend in (E) applies to (F,G,H). (G) Cell breadth and height, and (H) cell layer thicknesses measured in each stratum of TENOM. Data are expressed as mean ± S.D., *p <0.01**, p <0.01, ***p <0.001 and ****p <0.0001 as analyzed by ordinary One-way ANOVA with Tukey's correction (n = 3 independent biopsy with at least 10 measurements taken from each image).

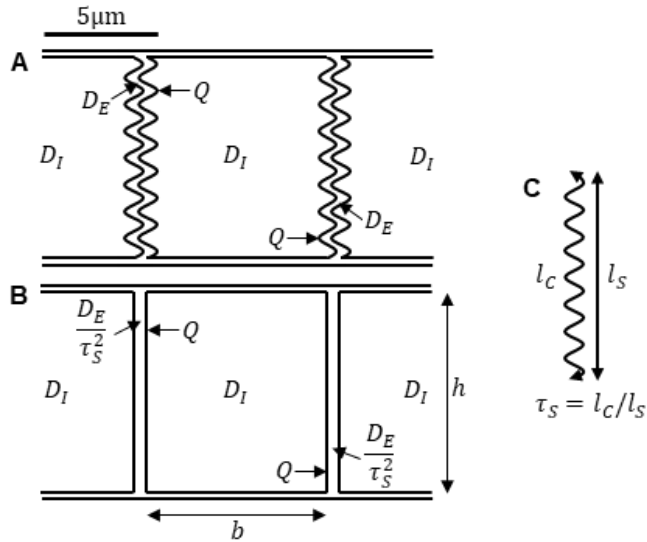

Supplemental figure 2. *In silico* schematic of chemical permeation through a normal oral mucosal (NOM) basal cell monolayer. Chemicals diffuse with coefficients $D_E$ (extracellular), $D_I$ (intracellular) and permeate cell membranes with rate $Q$ (A). (B) A straight-gap approximation has reduced extracellular diffusion. (C) Subcellular tortuosity $\tau_S > 1$ is the length of a convoluted gap $l_c$ relative to a straight gap $l_s$.

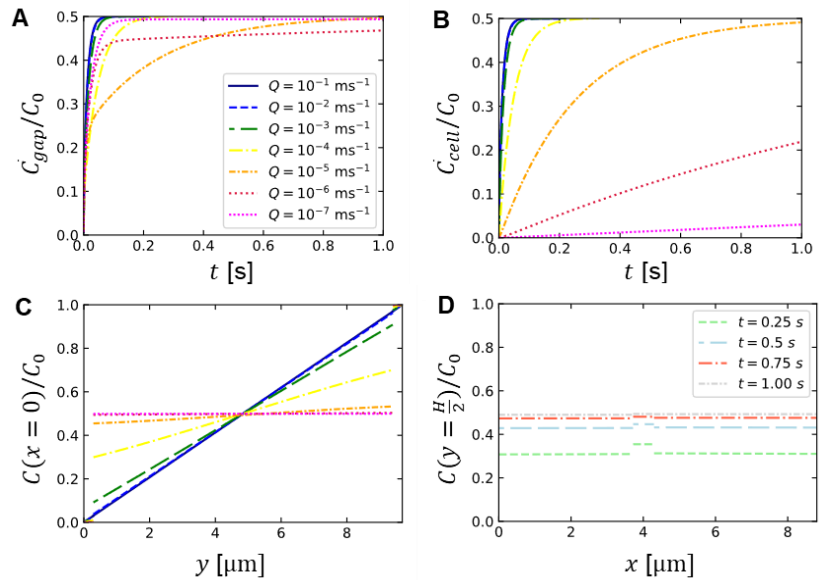

Supplemental figure 3. Effect of chemical property $Q$ on constant-supply chemical permeation through a monolayer of oral basal cells. A constant supply (concentration $C_0$) of a test chemical (diffusion coefficients $D_E = D_I = 7.5 \times 10^{-10} m^2 s^{-1}$) is applied at the apical cell surface and up taken below the monolayer. (A,B) Average drug concentration $\bar{C}$ in the central extracellular space (A) and intracellularly (B). The legend in (A) also applies to (B,C). Lipophilic chemicals $Q > 10^{-3}$ ms$^{-1}$ quickly reach a steady-state. Moderately lipophobic chemicals $10^{-6} < Q < 10^{-3}$ ms$^{-1}$ take much longer to reach steady-state. Chemicals which are highly lipophobic $Q < 10^{-6}$ms$^{-1}$ reach steady-state quickly in the gaps but slowly in the cells. (C) The concentration $C$ from the top to the bottom of the epithelium along the centre of cells (at $x = 0$) approaches a constant $C/C_0 = 1/2$ as the cell membrane permeability $Q$ is decreased. (D) Along the middle of the epithelium (on $y = H/2$) the difference in concentration between the central extracellular space and the cells is apparent at varying timepoints for $Q = 10^{-5}$ ms$^{-1}$.

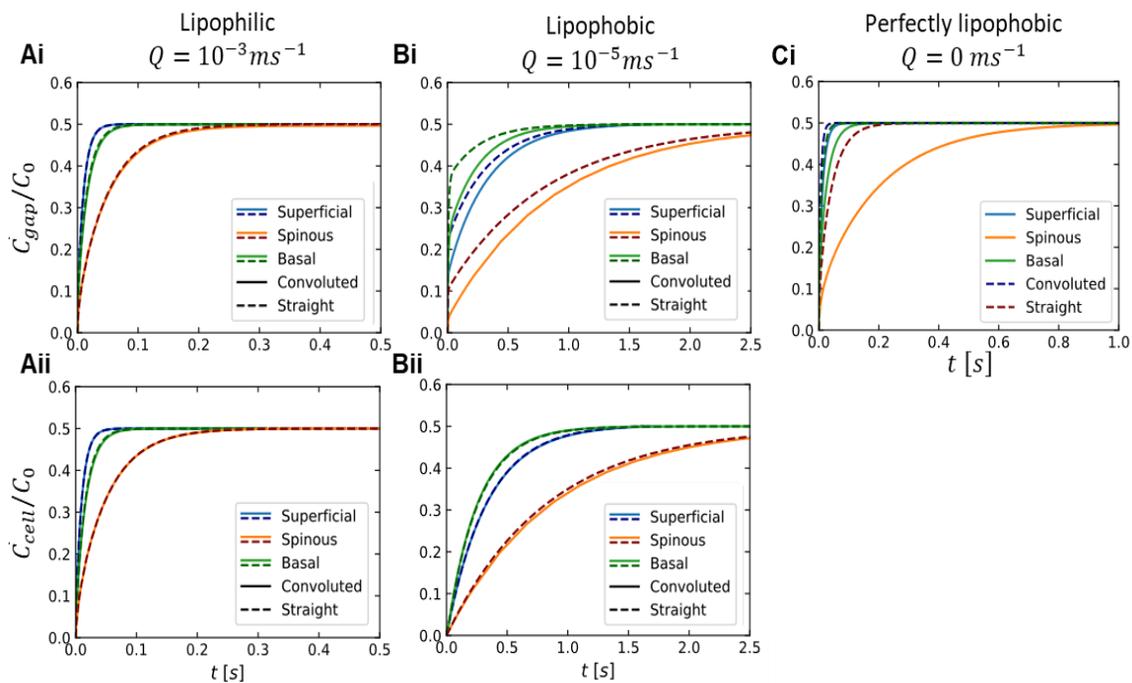

**Supplemental figure 4. Chemical permeation through a monolayer of oral keratinocytes modelled *in silico*: convoluted extracellular space compared to straight extracellular space without subcellular tortuosity accounted for $\tau_S = 1$.** Permeation through straight extracellular space is consistently quicker than in the physical convoluted model. However, by diminishing extracellular diffusion the straight-extracellular space model can provide a good approximation (see manuscript Fig. 2). Here a test chemical (diffusion coefficients $D_E = D_I = 7.5 \times 10^{-10}$m$^2$s$^{-1}$) permeates membranes with varying coefficients $Q$. A constant supply is applied with concentration $C_0$. Keratinocytes from each epithelial cell layer are modelled (superficial, spinous, basal) by incorporating physically relevant cell breadths, cell heights and convoluted extracellular space.